\documentclass[12pt,reqno]{amsart}
\usepackage{amsmath}
\usepackage{amssymb}
\usepackage{amsxtra}
\usepackage{amsthm}
\usepackage{color}
\usepackage{slashed}

\usepackage{hyperref}
\hypersetup{
    colorlinks = true,
    colorlinks = false,
    linkcolor = blue,
    anchorcolor = red,
    citecolor = blue,
    filecolor = red,
    pagecolor = red,
    urlcolor = blue
}
\usepackage{graphicx}

\usepackage{latexsym}
\usepackage{graphicx}

\def\ee{\end{equation}}
\def\bea{\begin{eqnarray}}
\def\eea{\end{eqnarray}}

\def\bs{\bar s}
\def\p{\partial}
\def\bp{\bar\partial}

\def\bb{\bar b}

\def\bR{\bar R}

\newcommand{\szego}{Szeg\"o\ }

\renewcommand{\Re}{{\operatorname{Re}\,}}
\renewcommand{\Im}{{\operatorname{Im}\,}}

\renewcommand{\epsilon}{\varepsilon}

\newcommand{\kahler}{K\"ahler }

\newcommand{\PP}{{\mathbb P}}

\newcommand{\R}{{\mathbb R}}
\newcommand{\C}{{\mathbb C}}

\newcommand{\Z}{{\mathbb Z}}

\newcommand{\kcalomega}{\mathcal{K}_{[\omega_0]}}

\newcommand{\CP}{\C\PP}

\newcommand{\dbar}{\bar\partial}
\newcommand{\ddbar}{\partial\dbar}

\newcommand{\E}{{\mathbf E}}

\newcommand{\bcal}{\mathcal{B}}

\newcommand{\dcal}{\mathcal{D}}

\newcommand{\hcal}{\mathcal{H}}

\newcommand{\kcal}{\mathcal{K}}

\newcommand{\mcal}{\mathcal{M}}

\newcommand{\ocal}{\mathcal{O}}
\newcommand{\pcal}{\mathcal{P}}

\newcommand{\De}{\Delta}

\newenvironment{rem}{\medskip\noindent{\it Remark:\/} }{\medskip}

\title{Random K\"ahler metrics}

\begin{document}

\author{Frank Ferrari,$^1$ Semyon Klevtsov$^{1,2}$ and Steve Zelditch$^3$}

\maketitle
{\small
\address{\it$^1$Service de Physique Th\'eorique et Math\'ematique, Universit\'e Libre de Bruxelles and}

\address{\it$\phantom{c}$International Solvay Institutes, Campus de la Plaine, CP 231, 1050 Bruxelles, Belgique}

\address{\it $^2$ITEP, B. Cheremushkinskaya 25, Moscow 117218, Russia}

\address{\it $^3$Department of Mathematics, Northwestern  University, Evanston, IL 60208, USA}
%\vspace{.1cm}
}

\begin{center}

\email{\tt\footnotesize frank.ferrari@ulb.ac.be, semyon.klevtsov@ulb.ac.be, zelditch@math.northwestern.edu}
\end{center}

\begin{abstract}

The purpose of this article is to propose a new method to define and calculate path integrals over metrics on a K\"ahler manifold. The main idea is to use finite dimensional spaces of Bergman metrics, as an approximation to the full space of \kahler metrics. We use the theory of large deviations to decide when a sequence of probability measures on the spaces of Bergman metrics tends to a limit measure on the space of all K\"ahler metrics. Several examples are considered.

\end{abstract}

%\date{August 8, 2008}

%\tableofcontents

\parskip 1ex

\section{Introduction}

\subsection{The aim of the paper and the basic setup}

This article is part of a series (cf.\ \cite{FKZ0,FKZ,FKZ1})  devoted to a new approach to   random  metrics on a manifold
$M$. It is assumed that $M$ is a \kahler manifold $(M, J, \omega_0)$, of any complex dimension $n$, with an
 integral \kahler form $\omega_0 \in H^2(M, \Z)$ and a compatible   complex structure $J$.   Motivation to study 
 random \kahler metrics arises in large  part from  the theory of two dimensional quantum gravity \cite{P}, because in this case any riemannian metric is K\"ahler for some complex structure. 
More generally, it is also of interest in both physics and mathematics  as a  theory of random surfaces and stochastic geometry in any complex dimension. In this article we introduce the general probabilistic and 
geometric foundations. Applications relevant to two dimensional quantum gravity are discussed in details in the companion articles
\cite{FKZ0,FKZ}.

The  novel  approach to defining
and studying random metrics  is based on some relatively new ideas in \kahler geometry involved in the Yau-Tian-Donaldson program of relating existence of constant scalar curvature \kahler metrics in the class of $\omega_0$
to stability in the sense of geometric invariant theory.  The main idea is to  approximate the infinite
dimensional space $\kcalomega$ of \kahler metrics in the cohomology class $[\omega_0]$  of $\omega_0$ by finite dimensional spaces $\bcal_k$ of 
``Bergman metrics," which are pullbacks of the Fubini-Study metric on complex projective space by Kodaira
type holomorphic embeddings defined by using holomorphic sections of powers $L^k$  of the line bundle $L$ with Chern class
$c_1(L) = [\omega_0]$.  Much current work in \kahler geometry involves the approximation of the infinite dimensional geometry of
$\kcalomega$ by the finite dimensional geometry of $\bcal_k$. The idea of this article is to use this approximation to regularize path integrals
over $\kcalomega$ by well-defined integrals over $\bcal_k$.
 
The space $\kcalomega$ can be parametrized by the 
\kahler potentials $\phi$ of  \kahler metrics $\omega_{\phi}$  in the cohomology class of  $\omega_0$,
\begin{equation}  \label{kcaldef} \kcal_{[\omega_0]} = \{\phi \in
C^{\infty}(M)/\R\; |\; \omega_{\phi}: = \omega_0 + i \ddbar \phi > 0\}
\end{equation}
with 
the equivalence relation that $\phi \equiv \phi'$ if and only if $\phi - \phi'$ is a constant. Since the riemannian metric associated to the \kahler form is  given by $ig_{a\bb} = \omega_{ a \bb}$, we use the same term 'metric' both for $g$ and corresponding \kahler form $\omega$.  We assume that the background \kahler form $\omega_0$ is integral, i.e.\ $\omega_0 \in H^{1,1}(M, 2 \pi \Z)$. In particular, if there exists a constant scalar curvature metric in the class $[\omega_0]$, it is natural to choose 
$\omega_0$ to be that metric.
In the case of a Riemann surface, one usually  parametrizes riemannian  metrics in a conformal class $\mbox{Conf}_{g_0}$ by a Weyl factor $\sigma$, with $g = e^{2\sigma} g_0$ for a background metric $g_0$. The cohomology class of $\omega_{0}$ is entirely determined by the area, and parametrizing the metrics by the K\"ahler potential $\phi$ instead of $\sigma$ is a simple change of variables given by
$e^{2\sigma} = 1 - \frac12\Delta_0 \phi$, where $\Delta_{0}$ is the positive laplacian for the metric $g_{0}$.
We refer to \cite{MW} for a detailed comparison of the relations between
$\mbox{Conf}_{g_0}$ and $\kcalomega$.  

Our aim is to define  probability measures on $\kcalomega$ and to study the geometry and physics
of the  resulting random K\"ahler manifolds. The natural action functionals
in this context are those arising in \kahler geometry. For example, the Aubin-Yau and Mabuchi actions, which are local in the K\"ahler potential $\phi$, are singled out, even though in complex dimension one they are non-local in the Liouville field $\sigma$. In \cite{FKZ0,FKZ} we discuss in detail an important r\^ole these new actions play in the problem of  the coupling of non-conformal quantum field theories to gravity.

Our approach relies on some recent developments in \kahler geometry. The  space $\kcalomega$ is formally an infinite
dimensional symmetric space of non-positive curvature, when
equipped with a riemannian metric introduced by Mabuchi \cite{M},
Semmes \cite{S2} and Donaldson \cite{Don1}. In a very strong sense, $\kcalomega$  is the infinite dimensional limit of the finite dimensional symmetric spaces $\bcal_k$ of Bergman
metrics. To be precise, $\bcal_k$ can be identified with positive-definite hermitian matrices of rank $N_k$ and
determinant one, i.e.\ with the coset $SL(N_k, \C)/SU(N_k)$. Here $N_k$ is the dimension of the space $H^0(M, L^k)$ of holomorphic sections of $L^k$.  The approximation of the differential geometry of  $\kcalomega$ by
that of $\bcal_k$ is now a fast developing theory, see e.g.\ \cite{Don,Don2,PS1,PS2,PS3,PS4,SoZ2,RZ,F}. We take this
theory in a new direction by using integrals over $\bcal_k$ to
regularize, or rigorously define, integrals over $\kcalomega$. At
the same time, the identification $\bcal_k\simeq SL(N_k, \C)/SU(N_k)$ makes an interesting direct link between random metrics and random
matrix models. Let us emphasize that these matrix models are of a novel type and are \emph{not} directly related to the matrix models used to study two-dimensional gravity in \cite{BrK,DS,GM,BZ,DGZ}.

\subsection{Defining path integrals}

In physics, path integrals must be defined by suitably regularizing and renormalizing formal infinite dimensional measures $e^{-
S(g)} \dcal g$. The main idea of this article is to define and compute such path integrals over metrics by finding a sequence of measures $\mu_k=e^{- S_k(g)} \dcal_k
g$ on the finite dimensional space $\bcal_k$ which tends to $e^{- S(g)} \dcal g$ in the sense that
\begin{equation} \label{INT1} \int_{\kcal_{[\omega_0]}} F(g) e^{- S(g)} \dcal
g := \lim_{k \to \infty} \int_{\bcal_k} F_k(g) e^{- S_k(g)}
\dcal_k g
\end{equation}
for suitable functionals $F$ on $\kcalomega$. The parameter $k$, or equivalently $N_{k}$, thus plays the r\^ole of the UV regulator.

Although we do not put the emphasis on applications to quantum gravity in the present work, let us mention that in this case a fundamental constraint on the measure $e^{-S(g)} \dcal g$ is background independence, i.e.\ independence with respect to the choice of $\omega_{0}$ in our context. A remarkable feature of the regularization scheme \eqref{INT1} is that the spaces $\bcal_k$ possess manifestly background independent measures $\dcal_k g$ for finite $k$, which converge to an infinite-dimensional background independent measure $\dcal\phi$. This is used in the companion papers \cite{FKZ0,FKZ,FKZ1}.

There are two natural ways to obtain measures on the space $\kcalomega$ of \kahler metrics. In the top-down approach, we start with an action functional $S$ of the space $\kcalomega$, which is suggested by a particular problem. For example, we show in \cite{FKZ0,FKZ} that the Mabuchi action enters naturally in the quantization of a non-conformal matter/gravity system. We then need to determine the approximating (regularized)
actions $S_k$. The simplest approach is just to restrict $S$ to the Bergman metrics $\bcal_k \subset \kcalomega$.  But there may be simpler and more natural approximations. For example, in  \S \ref{CP}, we propose an approximation scheme based on 
the so-called  {\it contraction principle} in the theory of large deviations. Let us note however that in general the finite dimensional approximations can be quite complicated and the corresponding large $k$ limit can be hard to study.

Therefore, it is also fruitful to consider a bottom-up approach where we choose sequences of measures $\mu_k  =e^{- S_k(g)} \dcal_k
g$  which are simple from the point of view of our formulation in terms of Bergman metrics, and study their 
large $k$ limits.  It is sometimes possible to determine the asymptotics of correlation functions of these sequences,
so they provide computable toy models for the general framework. 

However, in general, it is very difficult to determine asymptotics of correlation functions of the $\mu_k$. One can then focus on a simpler problem: does the sequence $\mu_k$ have an asymptotic
action, or in mathematical terms, does the
sequence $\mu_k$ satisfy a `large deviations principle' (LDP)?
The rate function of the LDP is the asymptotic action.   When we construct the sequence $\mu_k$ from
the bottom-up approach, we usually do not know in advance what infinite dimensional theory it tends to.
When the sequence satisfies an LDP with a rate function $S$,  we know that it tends in  weak sense to a theory
with action $S$.

The simplest sequences $\mu_k$ are those which are invariant under the unitary group $SU(N_k)$ which acts on $\bcal_k\simeq SL(N_k, \C)/SU(N_k)$ on the left. Such measures can be reduced to standard integrals over eigenvalues. As explained in \S \ref{EIGMODELS}, the $U(N)$-invariance of such measures and the reduction to eigenvalue integrals correspond to an infinite dimensional symmetry of $\kcalomega$ and the quotient by the symmetry is the space $\mcal(\R)$ of probability measures on $\R$. This reduction is a new observation in \kahler geometry and is related to the study of
geodesics of $\kcalomega$ \cite{RZ}. In \S \ref{Eigenvalue}, we discuss two sequences $\mu_k$ of this type. Note that in all cases, $SU(N_k)$-invariance leads to background-dependent
limits, typically to measures which are concentrated at or around the background metric.

\subsection{A few examples of interesting questions}

Once a notion of `random metric' is defined, one would ideally like to sample the ensemble of random metrics and graph the typical random manifold. 
Some natural geometric questions are: what is the `shape' of a random
manifold? How smooth is it? What is the expected volume
of a given subset $U \subset M$? What is the expected distance
between points or the expected diameter of the manifold?

These questions are often related to the computation of correlation
functions of natural observables on $\bcal_k$, such as values of the \kahler potential $\phi(z)$
or its derivatives at several points (see \S \ref{DEF}). 
Hence we would like to find non-trivial models for which these correlation functions are tractable.

A related circle of problems surrounds scaling relations of random metrics and the KPZ formula \cite{KPZ}, recently proved rigorously in \cite{DuSh}, see also the treatment in \cite{DB}. In the Liouville theory the KPZ relation can be derived by looking at the scaling of the path integral with the area $A$ of the surface \cite{Dav,DK}. A similar problem in our context is to understand how the Mabuchi path integral
\begin{equation}\label{MPI}
Z_k(A, \gamma) = \int_{\bcal_k(A)} e^{- \gamma S_M(\omega_0, \phi)} \dcal_k \phi
\end{equation}
behaves in the parameters $k, A$ and $\gamma$, where $A$ is the area of the surface and $\bcal_{k}(A)$ is the space of Bergman metrics 
of area $A$. Convergence of this integral is related to the ``stability" of $(M, L, [\omega_0], J)$, i.e.\ to the growth of the Mabuchi action $S_M(\omega_0, \phi) $ along geodesics of $\kcalomega$.

We plan to study these questions with the formalism introduced in the present paper in the future.

\subsection{Plan of the paper}

The paper is organized as follows. In \S \ref{Gb} we recall the geometric background. In \S \ref{DEF}, we explain our approach in detail. In subsequent sections, we consider different types of examples where the methods apply. We define probability measures on $\kcalomega$ using sequences $\{\mu_k\}$ of probability measures on $\bcal_k$.  We construct many interesting sequences $\mu_k$ which `converge' to an infinite dimensional limit in the sense that they satisfy a large deviations principle (LDP). Starting from a desired action functional $S$, we explain how to construct sequences $\mu_k$ satisfying an LDP with rate function $S$, by applying the contraction principle.  In \S \ref{Eigenvalue}  we study examples of sequences $\mu_k$ arising naturally form the identification of $\bcal_k$ as a coset space.  In the final section we build finite-dimensional approximations to geometric functionals using the contraction principle and pointwise convergence formulas.

\section{Geometric background}
\label{Gb}

\subsection{On the geometry of the space $\kcalomega$}

There is a natural riemannian metric
on the space $\kcalomega$ due to Donaldson, Semmes and Mabuchi \cite{Don1,S2,M} defined by 
\begin{equation}\label{MSD}  
G_\phi(\delta \phi_1, \delta \phi_2) = \int_M
\delta \phi_1(z) \delta \phi_2(z) \omega_\phi^{n}\, ,\quad \delta
\phi_1, \delta\phi_2 \in T_{\phi} \hcal \simeq C^{\infty}(M)\, .
\end{equation}  
It is shown in \cite{M,S2,Don1} that $G_{\phi}$ is of constant
sectional curvature, i.e.\ that $(\kcalomega, g_{\phi})$ is a
symmetric space. It is formally given by $\kcal_{[\omega_0]}
 = G_{\C}/G$ where  $G$ is the group of Hamiltonian symplectic
diffeomorphisms of $(M, \omega_0)$ and where $G_{\C}$ is the
complexification of $G$. The associated 
formal riemannian volume measure $\mathcal D\phi$ of
$\kcal_{[\omega_0]}$ is thus the  infinite dimensional
$G_{\C}$-invariant Haar measure. 

The infinite dimensional geometry is quite formal and many
features of finite dimensional symmetric spaces do not exist for
$\kcalomega$. For instance, $G_{\C}$ does not exist as a group \cite{Don1}, although it 
 is formally the group $\mbox{Diff}^{1,1}$ of diffeomorphisms $f$  of $M$ so that
$f^* \omega_0$ is of type $(1,1)$ with respect to the fixed complex structure $J$. Moreover, 
the exponential map of $\kcalomega$ is not well-defined \cite{RZ}, and in
directions where it is defined it is geodesically incomplete.
However, the situation is saved to a large extent by the existence
of the finite dimensional approximations $\bcal_k$  to
$\kcal_{[\omega_0]}$ that we describe in details later in this section.

Let us note that in the case of surfaces, where it is also natural to use the conformal Weyl factor $\sigma$ instead of $\phi$ to parametrize the metrics, one can consider the metric
\begin{equation}
\label{dewitt}
G_\sigma(\delta \sigma_1, \delta \sigma_2) = \int_M
\delta \sigma_1(z)\delta \sigma_2(z)
e^{2\sigma}\omega_{0}\, .
\end{equation} 
This metric and its associated formal integration volume form
$\dcal \sigma$ is induced from the well-known manifestly diffeomorphism invariant DeWitt-Ebin metric on the space of all riemannian metrics \cite{DeWitt,Ebin}. In terms of the K\"ahler potential, \eqref{dewitt} becomes the Calabi metric
\begin{equation}\label{Calabi}  G_\sigma(\delta \phi_1, \delta \phi_2) = \int_M
(\Delta_{\phi} \delta \phi_1(z)) (\Delta_{\phi}  \delta \phi_2(z))
\omega_{\phi},
\end{equation}
where $\Delta_{\phi}$ is the laplacian in the \kahler metric
$\omega_{\phi}$. Since it requires a Green's function to recover
$\delta \phi$ from $\Delta_{\phi}( \delta \phi)$, the two metrics \eqref{MSD} and \eqref{dewitt}
have quite different geometric properties. In particular, when endowed with the metric \eqref{MSD} the space $\kcalomega$ is negatively curved \cite{Don1,S2} and when endowed with the metric \eqref{Calabi} it is positively curved \cite{calamai,CR}.

\subsection{Random metrics and random matrices}

Crucial to our approach is the existence of finite dimensional approximations $\mathcal B_{k}$ to $\mathcal K_{[\omega_{0}]}$.
The properties of these approximations arise
from geometric \kahler quantization. Under the integrality
assumption $[\omega_0] \in H^2(M, 2 \pi \Z)$, there exists a
holomorphic line bundle $L \to M$ with $c_1(L) =[\omega_0]$. The space $\kcal_{[\omega_0]}$ is thus the same as
the space $\hcal$ of hermitian metrics $h$  on $L$ with curvature
$(1,1)$ form $\omega_h = -i\ddbar \log h \in \kcal_{[\omega_0]}. $
and we may identify \kahler metric $\omega_{\phi} \in
\kcal_{[\omega_0]}$ with the hermitian metric $h = e^{-
\phi}h_0$ or equivalently with the potential $\phi$. The \text{Hilb}ert
spaces `quantizing' $(M, \omega)$ are defined in the standard way
to be the spaces $H^0(M,L^k)$ of holomorphic sections $s_\alpha,$ 
$\alpha=1,...,N_k$ of
$L^k=L\otimes\cdots\otimes L$. We set the number of independent sections to be $N_k = \dim H^0(M, L^k)$.

The Kodaira embedding theorem uses a basis of these
holomorphic sections to embed $M$ into a high dimensional
projective space
$$z \to [s_1(z), \dots, s_{N_k}(z)]: M \to
\CP^{N_k - 1}. $$
The projective space has  a natural Fubini-Study metric on it. Its pull-back to $M$ is called a Bergman metric on $M$, and it depends on the choice of basis of sections we started with. Any two bases can be related by a $GL(N_k,\mathbb C)$ transformation and the space $\bcal_k$ can be identified with the space of all possible bases of sections which lead to inequivalent Bergman metrics. Since the Fubini-Study metric is invariant under $U(N_k)$ transformations of sections and since the overall normalization of the basis vectors is irrelevant since it does not change the associated Bergman metric, we have
\begin{equation}
\label{bk1}
\bcal_k \simeq SL(N_k, \C)/SU(N_k).
\end{equation}
The coset on the right side is the space of positive-definite hermitian matrices of determinant one.

This identification may be viewed in a slightly different way. We can identify $\bcal_k$ with the space $\hcal_k$ of all hermitian inner products on $H^0(M, L^k)$. The inner product associated with a given hermitian metric $h$ on $L$, which induces a hermitian metric $h^{k}$ on $L^{k}$, is given by
\begin{equation}
\label{Hilbmap}
{\text{Hilb}_k(h)}_{\alpha\beta}:=  \frac1V\int_M \bs_\alpha s_\beta h^k
\omega_{h}^n\, ,
\end{equation}
where $V$ is the volume, or in two dimensions the area $V=A$, of the manifold $M$. Having fixed a background metric $\omega_0 \in \kcal_{[\omega_0]}$, we denote the associated hermitian metric on $L$ by  $h_0 \in \hcal$.
This $h_0$ thus induces a reference inner product $\text{Hilb}_k(h_0)$, which we can fix, without loss of generality, to be equal to the identity matrix. Any other inner product may then
be identified with the positive hermitian
matrix representing it relative to the background inner product.  Since for any metric $h=h_0e^{-\phi}$ in $\hcal$ there is corresponding metric $\omega_\phi\in\kcalomega$, we thus obtain a map
\begin{equation} 
\label{HilbK} 
\text{Hilb}_k(\phi): \kcalomega \to \hcal_k:= \bigl\{\text{Inner products on}\ H^0(M, L^k)\bigr\}\, .
\end{equation}

We can also introduce a map $FS_k$ which converts an inner product
$G \in \hcal_k$ to a Bergman metric in $\bcal_k\in\kcalomega$ as follows. Given an inner
product $G $, one can form a $G$-orthonormal basis
$\{s_\alpha\}$ and an associated Kodaira holomorphic embedding.  If $e$ denotes a local frame for $L \to U
\subset M$, then any $s \in H^0(M, L^k)$ may be expressed locally
as $s = f e^{\otimes k}$.  The  pull back of  the Fubini-Study
hermitian metric on $\ocal(1) \to \CP^{N_k - 1}$ is given by
\begin{equation} \label{FSK} FS_k: \hcal_k \to \kcalomega\, , \quad FS_k(G) = \frac{1}{k} \log \sum_{j = 1}^{N_k}
|f_j(z)|^2 \in \bcal_k\subset\kcalomega. \end{equation} Given the background inner
product $\text{Hilb}_k(h_0)$, we let $s^0: M \to \CP^{N_k - 1}$
denote the associated Kodaira embedding.  Let $A \in GL(N_k, \C)$ denote the change of basis matrix from a $\text{Hilb}_k(h_0)$-orthonormal basis to a $G$-orthonormal basis $s$.  Then the positive matrix $P$ corresponding to $G$ is $P = A^\dagger A$ and the \kahler potential for the corresponding Bergman metric is given by
\begin{equation} \label{phiP} \phi_P(z) = FS_k(P)(z): =  \frac{1}{k} \log |A
s^0(z)|^2. \end{equation}
This formula is obviously invariant under the $U(N_k)$ action on $A$ on the left. Since also the space of potentials is defined modulo constant shifts, we thus have an identification
\begin{equation}
\label{FSID}
FS_k: SL(N_k,
\C)/SU(N_k) \to \bcal_k. 
\end{equation}

This identification can be viewed as
a correspondence between random positive ermitian matrices and
random Bergman metrics. It allows us to reduce the
asymptotic calculation of integrals over metrics to matrix
integrals. Thus, under the change of variables $P \in SL(N_k,
\C)/SU(N_k) \to \phi_P \in \bcal_k$, we may rewrite
integrals over metrics as integrals over matrices
\begin{equation} 
\label{INTEGRALS} \int_{\bcal_k} e^{- S_k(\phi)} F_k(\phi)  \dcal_k \phi =
\int_{SL(N_k, \C)/SU(N_k)} e^{- S_k (P) }F_k(P) d P\, .
\end{equation}
Different choices of the discretized action $S_k(P) = S_k(\phi_P)$, the operators $F(P)=F_k(\phi_P)$ and the measure $dP=\dcal_k \phi_P$ will be considered later in the text.

\begin{rem} We now have three ways to view the space $\bcal_k$: 

\begin{enumerate}

\item As a subspace $\bcal_k \subset \kcalomega$ of Bergman metrics in the space of \kahler metrics;

\item As the space $\hcal_k$  of inner products on $H^0(M, L^k)$;

\item As the coset space $SL(N_k, \C)/ SU(N_k)$. 

\end{enumerate}

The identification of (3) with (2) is just the choice of a background inner product, while the identification
of (1) and (2) uses the $FS_k$ map \eqref{FSID}. 

\end{rem}

Any \kahler metric on $M$ can be approximated by a Bergman metric with increasing accuracy as $k \to \infty$. This theorem follows from the large $k$ limit of the asymptotics of Bergman or \szego kernels \cite{Ti,C,Z}.
A key statement is that for any $\phi \in \kcalomega$,
\begin{equation} \label{FSHilb} \lim_{k \to \infty}  FS_k \circ \text{Hilb}_k(\phi) =
\phi.
\end{equation}
This limit formula may be differentiated any number of times.
There is even a complete asymptotic expansion for the left side in
terms of curvature invariants of the right side, which follows from the following result. If $\{s_{\alpha}\}$ is an orthonormal basis with respect to the inner product \eqref{Hilbmap}, then 
\begin{equation}
\label{TYZ}
\rho_k=\sum_{\alpha=1}^{N_k}\bs_\alpha(z)s_\alpha(z)h^k=k^n+\frac12k^{n-1}R(\omega_h)+\mathcal O(k^{n-2}),
\end{equation}
where for large $k$ there is a complete asymptotic expansion on the right, depending on various curvature invariants \cite{Z,C,Lu}, see also \cite{DKl}.  The left side of \eqref{TYZ} is the contraction of the diagonal of the \szego projector
\begin{equation} \label{PI} \Pi_{h^k}(z,w) = \sum_{\alpha} s_{\alpha}(z) \otimes s_{\alpha}(w)^*,  \end{equation}
i.e.\ the kernel of the orthogonal projection from all square integrable sections of $L^k \to M$ to the space $H^0(M, L^k)$
of holomorphic sections, with respect to the inner product  
\eqref{Hilbmap}.  The pointwise asymptotics \eqref{TYZ} provides an important tool in the problem of approximating the formal integrals over $\kcalomega$ by the integrals
\eqref{INTEGRALS}  over $\bcal_k$. However, as well as in random matrix theory, it requires more than just pointwise limits of integrands to obtain the $k\to\infty$ limits of matrix integrals.

\subsection{Invariant metric and the Mabuchi-Semmes-Donaldson metric}

Recall some facts on the group $GL(N,\C)$ and its subgroup $SL(N,\C)$. The group action on the hermitian matrices is defined as 
$$A \cdot P = A^\dagger P A\, . $$
Any matrix $g\in GL(N, \C)$ may be expressed uniquely as  the
product $g = k a n$ of a matrix   $k \in U(N)$,  $a \in
D(N)$, the subgroup of diagonal matrices and  $n \in T(N)$,  the
group of upper triangular matrices with ones on the diagonal. 
This factorization is known as
the Iwasawa decomposition of $GL(N, \C)$.
We denote the Lie algebras of $D(N)$ and $T(N)$ by
${\bf d}$ and $\bf t$ respectively. An important point is that
the exponential map  
\begin{equation} \label{EXPO}\exp_I: {\bf d} \oplus {\bf t} \to GL(N,\C)/U(N)  \end{equation} is a
diffeomorphism since $GL(N,\C)/U(N)$ has non-positive curvature. Hence, we may
define the inverse map $\log P$ of $P \in GL(N,\C)/U(N) $. In the case of
positive-definite hermitian matrices, it is the usual logarithm. 

The $GL(N,\C)$-invariant metric on the space $GL(N,\C)/U(N)$ can be written as
$\text{tr} (P^{-1}\delta P)^2$. With a suitable normalization, the corresponding Haar measure is 
\begin{equation} \label{HAAR} \mu_N(P) = \frac{1}{(\det
P)^{N}} d P_{11} \cdots dP_{NN} \prod_{1\leq j < k\leq N} d \Re
P_{jk}\, d \Im P_{j k}\, .
\end{equation}
Here we note that the space of positive-definite
hermitian matrices is open in the space of all hermitian matrices
and has $N$ real diagonal coordinates $P_{11} \cdots P_{NN}$ and $N(N - 1)$ off-diagonal coordinates $\Re P_{jk}$,  $\Im P_{j
k}$, for $1\leq j< k \leq N$, so that
the real dimension is $N^2$. The metric \eqref{HAAR} is easily pulled-back to the metric on $SL(N,\mathbb C)/SU(N)$ by imposing the constraint $\det P=1$.

Using the map \eqref{HilbK} one can show \cite{CS} that the invariant metric on $\bcal_k=SL(N_k,\C)/SU(N_k)$ converges to the Mabuchi-Semmes-Donaldson metric \eqref{MSD} on $\kcalomega$. Indeed, the variation of $\text{Hilb}_k$ is given by
\begin{equation}
\label{varHilb}
\delta \text{Hilb}_k(\phi)_{\alpha\beta}=\frac1V\int_M \left[(-k-\frac12\Delta_\phi) \bs_\alpha s_\beta h_0^ke^{-k\phi}\right]\delta\phi\,\omega_\phi^{n}\, .
\end{equation}
Then the invariant metric on $\bcal_k$ can be pulled back to a metric on $\kcalomega$ as follows
\begin{equation}
\label{HaarMab}
\text{tr} \bigl(\text{Hilb}_k^{-1}(\phi)\delta \text{Hilb}_k(\phi)\bigr)^2=k^{n+2}\int(\delta\phi)^2\omega_\phi^n+\mathcal O(k^{n+1}),
\end{equation}
where the right hand side follows from the expansion formula for the composition of Toeplitz operators \cite{MM}.
Thus, the discretized volume element $\mathcal D_k\phi$ in \eqref{INTEGRALS} corresponds to the Haar measure \eqref{HAAR}.

\subsection{Geometric action functionals}

The full measure on $\kcalomega$ will always be taken of the form $e^{-S}\mathcal D\phi$ for some action $S$. For example, in the case of surfaces, the Liouville action 
\begin{equation}\label{liouv}
S(g_0,\sigma)= \kappa S_L(g_0,\sigma)+\mu\int_M e^{2\sigma} \omega_{0}\, ,\quad S_L(g_0,\sigma)= \int_M \bigl( \p\sigma\wedge\bp\sigma + \sigma R_0 \omega_{0}\bigr)\, ,
\end{equation}
where $R_{0}$ is the Ricci scalar and $\mu$ the cosmological constant,
is a very natural choice. In applications to two dimensional quantum gravity, the positive coupling constant $\kappa$ is fixed in terms of the central charge of the matter CFT coupled to gravity. The Liouville action 
satisfies the cocycle identities,
\begin{equation}\label{cocyle} 
S_{L}(g_{0},g_{1})=-S_{L}(g_{1},g_{0})\, ,\quad
S_L(g_0,g_1)+S_L(g_1,g_2)=S_L(g_0,g_2)\, ,
\end{equation}
valid for any metrics $g_0$, $g_1$ and $g_2$ in the same conformal class. These identities are fundamental constraints that any effective gravitational action must satisfy. 

In our framework, it is natural to express the action functionals in terms of the K\"ahler potential, e.g.\ the Liouville action reads
\begin{equation}
S_L(\omega_0,\phi)=\int_M\biggl(\frac14\p\log\frac{\omega_\phi}{\omega_0}\wedge\bp\log\frac{\omega_\phi}{\omega_0}+\frac12R_0\omega_0\log\frac{\omega_\phi}{\omega_0}\biggr)\, .
\end{equation}
However, and quite remarkably, there are other geometric functionals, satisfying the cocycle conditions, that are local in $\phi$ (but not in $\sigma$ in the case of surfaces) that we can naturally consider. The most important one is the so-called Mabuchi action \cite{Mab} (often called ``Mabuchi energy'' in the mathematical literature), which reads
\begin{equation}\label{generalMab} S_{M}(\omega_{0},\phi) = \int_{M}
\biggl( \frac{\bar R}{n+1}\phi\sum_{k=0}^{n}\omega_{\phi}^{k}\wedge\omega_{0}^{n-k} - \phi\sum_{k=0}^{n-1} \text{Ric}(\omega_{0})\wedge\omega_{\phi}^{k}\wedge\omega_{0}^{n-1-k} + \omega_{\phi}^{n}\log\frac{\omega_{\phi}^{n}}{\omega_{0}^{n}}\biggr)\, ,
\end{equation}
where $\text{Ric}(\omega_{0})$ is the Ricci $(1,1)$-form associated with $\omega_{0}$ and $\bar R$ the average Ricci scalar given by 
$\bR=\frac{1}{V}\int_M R \omega_{\phi}^{n}$. The equation of motion of the Mabuchi action yields the condition for the scalar curvature to be constant, a property shared with the Liouville action restricted to a fixed K\"ahler class in one complex dimension. Another interesting functional, satisfying the cocycle identities, is the so-called Aubin-Yau action,
\begin{equation}\label{AYgen} S_{AY}(\omega_0,\phi) =\frac1{n+1} \int_{M} \phi\sum_{k=0}^{n} \omega_{\phi}^{k}\wedge\omega_0^{n-k}\, .\ee

In one complex dimension, the above formulas simplify and yields
\begin{align}
\label{AY}
& S_{AY}(\omega_0,\phi) =\int\biggl( \frac12\phi\p\bp\phi+\phi\omega_0\biggr)\, ,\\ \label{mabuchi}
& S_M(\omega_0,\phi) =\int\biggl(
\frac12\bR\phi\p\bp\phi+\phi\bigl(\bR\omega_0-\mbox{Ric}(\omega_0)\bigr)+\omega_\phi\log\frac{\omega_\phi}{\omega_0}\biggr)\, .
\end{align}
Notably, in this case the first two terms of the Mabuchi action are proportional to the Aubin-Yau action. 

\section{\label{DEF}Defining path integrals} 

\subsection{Formal definition}

The aim of this section is to explain how to define path integrals of the form $\int e^{- S} \dcal
\phi $, where $S$ is a desired action and $\dcal \phi$ is
the volume form of the Mabuchi-Semmes-Donaldson riemannian metric.
The infinite dimensional, continuum path integral is only a formal expression. We would like to
define path integrals over $\kcalomega$ as a limit $k\to\infty$ of the finite-dimensional, regularized, path integrals over a sequence of
measures $\mu_k=e^{- S_k(\phi)} \dcal_k \phi$ on $\bcal_k$:  
\begin{equation}
\label{INT} \int_{\kcal_{[\omega_0]}} F(\phi) e^{- S(\phi)} \dcal
\phi := \lim_{k \to \infty} \int_{\bcal_k} F_k(\phi) e^{- S_k(\phi)}
\dcal_k \phi\, .  \end{equation}
To define rigorously the limit on the right side of this equation is clearly a difficult issue, related to the renormalization of the theory, a problem that we will only start to address in the present work.

Let us comment on the choice of the volume forms on $\bcal_k$ and $\kcalomega$. As we explained in \S \ref{Gb}, the
obvious choice is to take  $\dcal_k \phi$ to be the Haar measure
on $\bcal_k$ \eqref{HAAR}, i.e.\ the volume form of the symmetric space
riemannian metric, which corresponds in the infinite dimensional limit  to the riemannian volume density associated
to the metric (\ref{MSD}). Another option is to use the fact  that the exponential map \eqref{EXPO} $\exp_I: T_I \bcal_k \to \bcal_k$
is a diffeomorphism, and that the tangent
space is equipped with the natural Euclidean volume form. This volume form converges to the tangent space volume form on $\kcalomega$, which is analogous to the flat volume form used in the Liouville theory \cite{Dav,DK}. Here we will consider both options.

The main problem is then to construct sequences $S_k$ of finite dimensional actions, so that the right side of Eq.\ (\ref{INT}) converges to a finite theory with the desired action $S$ on $\kcalomega$. This is the problem of renormalization. Mathematically, we need to explain
what it means for a sequence of measures $\mu_k$ on $\bcal_k$ to converge
to a measure on $\kcalomega$.  Since $\bcal_k \subset \kcalomega$ we may regard the $\mu_k$ as a sequence of measures on $\kcalomega$ which are concentrated on $\bcal_k$. Hence we can ask whether the sequence $\mu_k$ has a weak limit in the sense of probability theory, i.e.\ whether the right side of \eqref{INT} converges for some functions $F$. 

A natural class of correlation functions $\kcal_k(z_1, \dots, z_n)$ is given by taking $F$ in Eq.\ \eqref{INT} to be the product of the \kahler potential field $\phi$ at several points,
\begin{align} \nonumber  \kcal_k(z_1, \dots, z_n)  & = 
 \int_{\bcal_k}  \prod_{j = 1}^n (\phi(z_j))  e^{- S_k(\phi)} \dcal_k \phi \\ \label{GCF} & =  \int_{SL(N_k)/SU(N_k)} \left( \prod_{j = 1}^n \log 
\bs(z_j)Ps(z_j) \right)  e^{- S_k(P)}  \mathcal D_{k}P\, ,  \end{align} 
where in the second line
we use \eqref{phiP} and \eqref{INTEGRALS}  to change variables from the 
K\"ahler potential $\phi$ to the positive-definite hermitian matrix $P$ and $\mathcal D_{k}P$ is an appropriate measure. In some simple examples,
$e^{- S_k(P)} \mathcal D_{k}P $ splits into the Haar measure on $U(N_k)$ and the measure on the eigenvalues of $P$, in which case it is often possible to calculate some of these correlation functions at large $k$.

\subsection{\label{LD}Large deviations approach}

A simpler criterion for convergence of a sequence of finite dimensional measures $\mu_k = e^{- S_k} \dcal_k \phi$ to the infinite dimensional measure $e^{- S} \dcal \phi$  is in terms
of large deviations theory. In fact, a sequence may satisfy an LDP without converging in the previous sense, i.e.\ without all correlators necessarily having finite limits, so the LDP may be viewed as a rough check on whether the sequence is defining an infinite dimensional theory.

As has been already mentioned, we
may regard the measures $\mu_k$ on $\bcal_k $ as
measures on $\kcal_{[\omega_0]} $ which are supported on (i.e.\
are zero away from) the submanifolds $\bcal_k\in\kcalomega$. One says that the
sequence of measures $\mu_k$ satisfies an LDP 
with the speed $n_k$ and a rate function $S$ if, for all balls
$B(\omega, \delta) \subset \kcal_{[\omega_0]}$ of radius $\delta$,
one has
\begin{equation} \label{LDP} \lim_{k \to \infty} \frac{1}{n_k} \log \mu_k (B(\omega, \delta)) = -
\inf_{\omega_{\phi} \in B(\omega, \delta)} S(\omega_{\phi}).
\end{equation} 
Here, the radius of the ball $B(\omega, \delta)$ is computed with the riemannian metric on $\kcalomega$. Thus, a basic
criterion for the sequence $\mu_k$ to tend to $e^{- S(\phi)} \dcal \phi$ is that it satisfies an LDP with the rate functional $S$. This criterion is
rather crude since it only determines the action $S$ and does not
take into account the integration measure $\dcal \phi$. But this
is an advantage when the infinite dimensional measure is hard to define and only the corresponding action is known.

It may be hard to compute the rate function by integrating over balls. However, one could use a simpler method, based on G\"artner-Ellis theorem, see \cite{DZ} for background.
Suppose that  there exists a limit logarithmic moment generating function
$$
W(J) =\lim_{k\to\infty} \frac{1}{n_k} \log \E_k \{e^{ \langle J,\phi_k\rangle} \}\, ,
$$
where $\phi_k$ is our random variable, and $\langle \,,\rangle $ is an appropriate pairing, which is case-dependent.
Then the LDP rate functional is determined by the Legendre transform of the 
generating function
$$
S(\phi)=\sup_{J}\{ \langle J,\phi \rangle-W(J)\}\, .
$$
This method is a finite-dimensional analog of the method of effective action in the quantum field theory, where the generating functional is defined  as a Fourier-Laplace transform of the full quantum measure. 

\subsection{\label{CP}Contraction principle}

No general geometric method for constructing simple and tractable approximating actions $S_k$ to a given geometric action $S$ is known. The most straightforward way is to restrict the functional $S$ to $\bcal_k$
\begin{equation}
\label{rest}
S_k = S(\phi) |_{\bcal_k}.  
\end{equation}
This is a good approximation in the sense that
$S_k\simeq S(\phi)$ as $k \to \infty$. On the other hand, the restriction does not simplify
the action nor does it arise naturally from the point of view of the geometry of $\bcal_k$. 

In special cases, considered in the last section, such as the Aubin-Yau action or
Mabuchi action, the pointwise Bergman kernel expansion
\eqref{TYZ} can be used to construct alternative finite dimensional
approximations. Other techniques related to the geometric
invariant theory have been used \cite{Zh,PS3,PS4,PS2}, see also  \cite{F}. These approximations are used to construct $S_k$ so that $S_k(T_k(\phi)) \to S(\phi)$, where
\begin{equation} \label{TK} T_k = FS_k \circ \text{Hilb}_k: \kcal_{[\omega_0]} \to \bcal_k\, .\end{equation} 
It is not clear whether knowing such pointwise approximations is enough to approximate the path integrals.

However, the contraction principle of  large deviation theory suggests some new methods for constructing finite dimensional approximations. Namely, we may define
\begin{equation} \label{SKPROP} 
S_k(P) = \inf_{\{\phi \in \kcal_{[\omega_0]},\, \text{Hilb}_k(\phi) = P\}} \;\;
S(\phi).  \end{equation}
It is clear that the $S_k$ of \eqref{SKPROP} does tend to $S$ as $k \to \infty$ since the fiber of the map $\text{Hilb}_k(\phi)$ shrinks to a single point $\{\phi\}$ in this limit. We now explain how \eqref{SKPROP} is related to the contraction principle.

 If we are given  a formal infinite
dimensional measure $\mu$,  we can form  approximating measures
by pushing $\mu$ forward to $\bcal_k$, meaning
\begin{equation} \label{PF} \mu_k = (\text{Hilb}_k)_* \mu\, , \end{equation}
where $(\text{Hilb}_k)_*$ denotes the push-forward under the map
(\ref{HilbK}). It is equivalent but often more natural
geometrically to consider instead the pushforward
\begin{equation} \label{PFTK} \mu_k = (T_k)_* \mu\, . \end{equation}
We observe that 
\begin{equation}\label{PUSHBACK} \lim_{k\to\infty}\mu_{k}= \mu\, , \end{equation} so that this method of obtaining finite
dimensional approximations is devised to automatically reproduce the original measure in the large $k$ limit. Indeed, the statement
(\ref{PUSHBACK}) is equivalent to
$$\lim_{k \to \infty} \int_{\kcalomega} f \bigl(FS_k\circ \text{Hilb}_k(\phi)\bigr) \mu_k = \int_{\kcalomega} f (\phi)
\mu$$ for suitable functions $f$. But this holds by
(\ref{FSHilb})  as long as the limit can be taken under the
integral sign since
$$\lim_{k \to \infty} f (FS_k \circ \text{Hilb}_k(\phi)) = f(\phi) $$
for all continuous functions $f$.

The contraction principle 
(see e.g.\ \cite{DZ,Gu}) says that given a sequence of measures $\mu_k$
satisfying an LDP with speed $n_k$ and  rate function $I$, and a map $T: X \to Y$, the pushforward measures $T_* \mu_k $ also satisfy an LDP with speed $n_k$  and the rate function 
\begin{equation}
\label{Contr}
J(y) = \inf_{\{x\in X,\,  T(x) = y\}} I(x)\, .
\end{equation}

Our proposal  \eqref{SKPROP}  does not in fact  follow from the contraction principle because we have only one measure $\mu$ on 
$\kcalomega$ rather than 
a sequence of measures and also because we have a sequence of maps $T_k$
between $X=\kcalomega$ and $Y=\bcal_k$ rather than one fixed map $T$, as in \eqref{Contr}. Therefore we view  \eqref{SKPROP} as an approximation scheme suggested by the 
contraction principle, and the formal considerations above only suggest why it is natural. 

This procedure is actually reminiscent of the Wilsonian notion
of renormalizing path integrals by integrating out higher
frequencies, in the sence that if we have $\mu_k \simeq e^{- n_k I (\phi)} \dcal (\phi)$ then the pushforward measure corresponds to computing the path integral $$T_* \mu_k
\simeq \int_{T^{-1}(\phi)}  e^{- n_k I (\phi)} \dcal (\phi)$$ over all metrics that correspond to the same regularized Bergman metric.

In principle, we can modify our problem so that the contraction principle does apply. 
For this we need to embed the  measure $\mu$ and action $S$ into a one parameter family $\mu_{\epsilon}, S_{\epsilon}$, satisfying an LDP. Then we can apply the map $T_k$ to a sequence $T_k{*} \mu_{\epsilon}$. We now have two parameters $k, \epsilon$.
We can then apply the contraction principle with $k$ fixed to the family in $\epsilon$. 

Since there are many ways to embed $\mu, S$ into a family $\mu_{\epsilon}$ satisfying an LDP, we would like to choose one which leads to a good approximation as $k \to \infty$. For some special $\mu, S$ there may be natural parameters one can use for $\epsilon$. We now consider two ways
to proceed when no such parameters occur. 

First,  if the measure $\mu$ on $\kcalomega$ is well-defined, 
 we can just replace our path integral $\int e^{- S(\phi)} \dcal \phi$
by $\int e^{- \frac{1}{\epsilon}  S(\phi)} \dcal \phi$. This new family automatically has
an LDP with rate function $S$ and it directly gives  \eqref{SKPROP}  as the approximate rate function.

The approximation  \eqref{SKPROP}  can often be difficult to evaluate since it is a constrained variational problem. 
A drastic simplification is to linearize it around the critical metric $\varphi_c$, in the case when the action $S(\varphi)$  has a unique  critical point. With no essential loss of generality we may assume $S(\varphi_c) = 0$.
We then change the action from $S$ to $S_{\epsilon}$ depending on  a small parameter $\epsilon$ by 
defining
$S_{\epsilon} (\varphi) = \frac{1}{\epsilon} S(\varphi_c+\sqrt\epsilon  \eta)$. We expand the right side in powers of $\epsilon$,
$$
\frac1\epsilon S(\varphi_c+\sqrt\epsilon \eta)= \frac12\int
\eta\frac{\delta^2 S}{\delta\varphi^2}|_{\varphi_c}\eta+\mathcal
O(\sqrt\epsilon).
$$
We then change variables $\phi_0 + \epsilon \eta \to \eta$ in the functional integral.  We are simply approximating
the original measure by a gaussian measure on the tangent space $T_{\phi_c} \kcalomega$. 
We ignore the Jacobian
factor of powers of $\epsilon$ since we also normalize the measure by dividing by its mass, which has the
same power of $\epsilon$. The resulting integral is then a gaussian integral 
$$\int_{ T_{\phi_c} \kcalomega}  e^{ - \frac{1}{\epsilon} \left[  \langle \delta^2 S_{\varphi_c} \eta, \eta \rangle+ \langle J, \eta \rangle \right]}
\dcal \eta $$which satisfies an LDP with rate function $\delta^2 S_{\varphi_c}. $  Hence we may apply the contraction principle
to this sequence (or family) of measures when we push forward by $$D\text{Hilb}_k(\phi_c): T_{\phi_c} \kcalomega \to 
T_{\text{Hilb}_k(\phi_c)} \bcal_k\, . $$  Here we use that
$\text{Hilb}_k(\phi_c + \sqrt{\epsilon} \eta) = \text{Hilb}_k(\phi) + \sqrt{\epsilon} D\text{Hilb}_k(\phi) \eta + O(\epsilon). $
Since the rate function for 
the family is the quadratic part $S_2(\eta) = \langle \delta^2_{\phi} S_{\varphi_c} \eta, \eta \rangle, $ the finite
dimensional approximation to this rate functional is given by
\begin{equation} \label{IK} S_k(\delta P) = \inf_{\eta: D \text{Hilb}_k(\phi_c)   \eta = \delta P} I_2(\eta)\, .  \end{equation}
The result we get is equivalent to the well-known fact that the pushforward of a gaussian measure under a linear
map is a new gaussian measure. 

This linearizes the problem to the tangent spaces and we would only expect \eqref{IK} to be a good approximation to $S$ near the critical point $\varphi_c$.

\section{Measures of the matrix model type on Bergman metrics}
\label{Eigenvalue}

In this section we consider the bottom-up approach, namely we take the finite dimensional integrals (\ref{INTEGRALS}), that arise naturally from the random matrix side of the identity rather than from the geometric side. In particular we consider $U(N_k)$-invariant measures. We emphasize that the corresponding actions $S_k(\phi)$ do not necessarily have a good geometric interpretation. In this approach geometric features of \kahler metrics are suppressed and only the spectral theory
of random positive-definite hermitian matrices is retained. Hence one should not expect to get a geometric theory of random surfaces arising in the large $k$ limit. Our purpose is simply to illustrate the LDP's for such measures, and to compute some interesting correlation functions.

\subsection{\label{EIGMODELS}Eigenvalue models}

From the point of view of the standard matrix models the simplest class of measures on $\bcal_k$ has the form 
\begin{equation}
\label{eig}
\mu(P)= \mathcal F(\Lambda) dP,
\end{equation}
where $\mathcal F(\Lambda)$ depends only on the eigenvalues $\Lambda$ of the matrix $P\in\bcal_k$, and not on the angular part $U$ in the decomposition $P= U\Lambda U^{\dagger}$ and the measure $dP$ can be decomposed as $d\Lambda d U$,
where $dU$ is the Haar measure on $U(N_k)$. However, we should emphasize that even for the eigenvalue type measures the computation of geometric correlation functions  \eqref{GCF} will involve nontrivial integration over angular variables.

In principle, the integration in \eqref{INTEGRALS} goes over the matrices $P\in\bcal_k$ which have unit determinant.
In the actual computations it is easier to use an unrestricted positive hermitian matrix $P\in GL(N_k)/U(N_k)$ as integration variable. Since the dilatations $P\to c P$ for $c>0$ do not change the associated Bergman metric, they correspond to a `gauge' freedom that must be mod out. We shall do that in the standard way by inserting the gauge constraint $\det P=1$ in the path integral with a $\delta$-function, as follows,
\begin{equation}
\label{constr}
\delta(\log\det P)\mu(P)=\int_{-i\infty}^{i\infty}da\, [\det P]^a\mu(P)=
\int_{-i\infty}^{i\infty}da\, \mu_a(P)\, .
\end{equation}
Obviously, the constrained measures are still of the type of Eq. \eqref{eig}. 

\subsection{Empirical measures} 

Recall that, after  fixing
a background metric $\omega_0$, any inner product $\text{Hilb}_k(\phi)$ may be represented as a positive hermitian operator $P_k(\phi)$  relative to $\text{Hilb}_k(\omega_0)$. In mathematical terms, we can encode \kahler potentials by the  empirical measure of eigenvalues (that is to say, by the density of eigenvalues) of either  $\frac{1}{k} \log P_k(\phi)$ or $P_k(\phi)$. In the first case we encode the hermitian operator  \begin{equation} \label{LK} X_k(\phi) : =
\frac{1}{k} \log P_k(\phi) \end{equation} 
by the density
\begin{equation} \label{EM} \nu_{X_k(\phi)}(x) = \frac{1}{N_k} \sum_{j = 1}^{N_k} \delta (x-{x_{k,j}})\, , \end{equation}
where $\{x_{k, j}\}_{j = 1}^{N_k}$ are the eigenvalues of
$X_k(\phi)$. This density defines a measure $d\nu_{X_k(\phi)} = 
\nu_{X_k(\phi)} dx \in \mcal(\R)$, where $\mcal(\R)$ is the convex set of probability measures on $\R$.
In the second case we can do the same for $P_k$. We can then choose our matrix model such that at large $k$ the density \eqref{EM} associated with the matrix $\frac{1}{k}\log P_{k}$, or the density associated with the matrix $P_{k}$ itself, has a smooth large $k$ limit. Clearly this would correspond to quite different models. But in either case we have an embedding
\begin{equation} \nu_k \circ \text{Hilb}_k : \kcal_{[\omega_0]} \to \mcal(\R_+). \end{equation}
Under this embedding, we may  push forward the given
probability measure on $\kcal_{[\omega_0]}$ to obtain a probability measure on
$\mcal(\R)$:
\begin{equation} (\nu_k \circ \text{Hilb}_k)_* \mu = {\bf Prob}_k \in \pcal(\mcal(\R)), \end{equation}
where $\pcal (\mcal(\R))$ is the metric space of probability
measures on $\mcal(\R_+)$. The rate function for
the sequence of measures ${\bf Prob}_k$, if it exists, will then be a functional on $\mcal(\R)$ rather than a geometric functional. This approach is  common in the theory of large deviations of random matrices, see e.g.\ \cite{Gu}, and comes up naturally in our problem.

We also remark that the empirical measure map  induces a new family of correlation functions, the spectral  correlation functions on $\R^n$,  defined by

\begin{equation} \label{SCF}\kcal_k(\mu_1, \dots, \mu_n) = \E_k
\prod_{j = 1}^n \nu_k(\mu_1) \otimes \cdots \otimes \nu_k(\mu_n).
\end{equation}
Here, $\nu_k \otimes \nu_k \cdots \otimes \nu_k$ denotes the
product  probability measure on $\R^n$ determined by $\nu_k$,  i.e. its integral against a function $f(\mu_1, \dots,
\mu_n)$ is given by $\int_{\R^n} f(\mu_1, \dots, \mu_n)
\nu_k(\mu_1) \cdots \nu_k(\mu_n). $

\subsection{Wishart ensemble}

The Wishart measure is defined by
$$\mu_a(P)=e^{- g \,\text{tr} P}[\det P]^{a+N}\mu_N(P). 
$$ 
Here
$\mu_N(P)$ is the Haar measure (\ref{HAAR}). Under the decomposition $P = U\Lambda U^{\dagger}$ the radial part of the Wishart measure is the eigenvalue distribution
\begin{equation}
\label{radial}
\frac{1}{Z_N} e^{- g  \sum_{j = 1}^N \lambda_j}
 |\Delta(\lambda)|^2 \prod_{j=1}^N \lambda_j^a \, d \lambda_j\, ,
\end{equation} 
where $\Delta(\lambda) = \prod_{j<k}(\lambda_{j}-\lambda_{k})$ is the usual Vandermonde determinant. The normalizing
constant is $Z_N=\int\mu(P) =\int_{-i\infty}^{i\infty} da\, g^{-N(N+a)}\prod_{j = 1}^N \Gamma(j + 1) \Gamma(j+a)$. We have also mod out by the angular integral.

Now, we would like to perform a large deviation analysis for the sequence of probability measures \eqref{radial}. Since we have $U(N)$ invariance, we only need to study the logarithmic asymptotics of integrals over eigenvalues, which is of a type studied in \cite{J} (see Theorem 2.1).
To obtain the large $N$ limit, it is necessary to embed the spaces $\R_+^{N}$
into one fixed space.  In the context of Wishart models, the density of eigenvalues
$$d\nu_{\lambda}(x): =
\frac{1}{N} \sum_{j = 1}^{N}  \delta(x- \lambda_j)\,dx $$ 
of the matrix $P$ will have a smooth limit. Positive
matrices with the same density of eigenvalues are conjugate under
$U(N)$. Thus we have mapped the symmetric space $\pcal_{N}$
into the space $\mcal_1^+(\R_+)$ of probability measures on the
positive real numbers $\R_+$. We then pushforward the measure $d\nu_N$
 under this map to define  a probability measure $P_N$ on $\mcal_1^+(\R_+)$.
Thus, $P_N$ is supported on the set of measures \eqref{radial} in terms
 of $d\nu_{\lambda}$ as
\begin{equation} \label{DENS}  |\De(\lambda)|^2  e^{ -  g\sum_{j=1}^N\lambda_j} \prod_{j = 1}^{N} \lambda_j^{a} =
e^{- gN \int_{\R_{+}} xd\nu_\lambda+ aN \int_{\R_{+}} \log x \,d\nu_{\lambda}
+ N^2 \Sigma(\nu_{\lambda})},
\end{equation} where  $$\Sigma(\nu) = \int_{\R_{+}} \int_{\R_{+}} \log |x - y| d\nu(x)\,
d\nu(y)$$ is the standard Coulomb repulsion term between eigenvalues. The coupling constant in \eqref{DENS} scales as $g=g_0N$, which is the usual 't~Hooft's scaling.
Then the sequence of measures \eqref{radial} satisfies a large deviation principle with the rate function  \cite{J}
\begin{equation} \label{WISHARTI} S(\nu_\lambda) = -g_0 \int_{\R_{+}} x d\nu_\lambda + \Sigma(\nu_{\lambda})\, . \end{equation}
Also, integration over $a$ imposes the constraint 
\begin{equation}\label{cons} \int_{\R_{+}}\log x \, d\nu_\lambda=0\, .\end{equation}
It is not difficult to solve the saddle point equation associated with \eqref{WISHARTI} taking into account the constraint \eqref{cons}. The result is a smooth eigenvalue distribution $\nu_{\lambda}$ that depends on $g$ and whose support is an interval of $\R_{+}$ containing the point $x=1$. Since the details of this solution will not be used presently, we shall explain it in a separate publication.

It is also possible to explicitly calculate the geometric
correlation functions \eqref{GCF}. Here we would like to quote the results which will appear elsewhere. We start with the choice of a basis of sections $s$, which is orthonormal with respect to a choice of background metric $h_0^k$ on the line bundle $L^k$, with corresponding background \kahler metric $\omega_0(z)=-\frac1ki\p\bp\log h_0^k$.
The random metric is then parametrized as $\omega_{\phi(P)}(z)=\frac1ki\p\bp\log\bs(z)Ps(z)$. 
The expectation value of a single metric in the large $k$ limit can then be easily shown to be equal to the background metric
$$
\E_k \omega_{\phi}(z) = \frac1ki\p\bp\log | s(z)|^2=\omega_0(z)+\mathcal O(1/k)\, .
$$
In fact, it is not hard to show that the same result holds for an arbitrary measure of the eigenvalue type \eqref{eig}
$$
\int_{\bcal_k} \omega_{\phi(P)}(z)\,\mu(P) =\omega_0(z)+\mathcal O(1/k)
$$
In this scaling limit the two-point function has a similar behavior
$$
\E_k \omega_{\phi}(z)\omega_{\phi}(y)=\omega_0(z)\omega_0(y)+\mathcal O(1/k)\, .
$$
Essentially, these calculations indicate  that in the
large $k$ limit the Wishart measures concentrate at the Dirac measure $\delta_{\omega_0}$, i.e.\ that random metrics concentrate at the
background metric. 

This result is not contradictory with the fact that the eigenvalue distribution of $P$ has a non-trivial smooth limit as indicated above. Indeed,  the eigenvalues of $P_{k}(\phi)$ for $\phi\not = 0$ grow (or decay) exponentially with $k$, in such a way that the eigenvalue distribution of  $X_k$ \eqref{LK}-\eqref{EM}
%
%\begin{equation}\label{Xkdef}
%X_{k}=\frac{1}{k}\log \text{Hilb}_{k}(\phi)
%\end{equation}
%
has a smooth limit. This scaling follows from the large $k$ asymptotics \eqref{HaarMab} and the fact that the size $N_{k}$ of the matrix grows as $k^{n}$. 

More precisely, let $\nu_{X_{k}(\phi)}(x)$ be the density of eigenvalues of $X_{k}(\phi)$ and let $\nu_{\phi}(x)$ be its limit when $k\rightarrow\infty$. We call $d\nu_{\phi}$ the {\it \szego limit measure} of
$\phi$.  We  conjecture  that the limit exists and  is given by
\begin{equation} \label{SZEGO} d\nu_{\phi} = (\exp_{\phi_0}^{-1} (\phi))_* d\nu_{\phi_0}, \end{equation}
where
$$\exp_{\phi_0}: T_{\phi_0} \kcal_{[\omega_0]} \to \kcal_{[\omega_0]} $$
is the exponential map for the Donaldson-Mabuchi-Semmes metric \eqref{MSD}. Here, 
$$\exp_{\phi_0}^{-1} (\phi) =  \dot{\phi}_0\, ,
$$
is the initial tangent vector to the unique
geodesic with endpoints $\phi_0, \phi$. 
 The conjecture follows formally from the fact that 
$\frac{1}{k} \log P_k(\phi)$ is the (scaled) inverse of the exponential map of $\hcal_k$ and that
geodesics of $\hcal_k$ tend to geodesics of $\kcalomega$ \cite{PS1}. The authors have verified
the conjecture in the case of toric varieties.

These considerations motivate the study of the model of the next subsection.

\subsection{\label{logP}Models on the tangent space}

Let us thus use $X_{k}$ defined by \eqref{LK} as the basic variable.  In effect, we are inverting the exponential map
\eqref{EXPO} and defining the probability measure on the tangent 
space $T_I \bcal_k$ of the symmetric space of positive
matrices.

The simplest such measure is the gaussian measure 
\begin{equation}\label{gaussmodel}
\mu(X)=e^{- gN\text{tr} X^\dagger X} d X\, . 
\end{equation}
Here $dX$ is the usual Lebesgue measure on the tangent space. The action $\text{tr} X^\dagger X$ is simply the square of the geodesic distance in the symmetric space from $0$ to $\exp_0 X$. Indeed, the
exponential map follows the unit speed  geodesic from $0$ to $\exp_0 X$ and its length is $||X||$. Consequently, the
action has the natural infinite dimensional limit 
$$S(\phi_0, \phi) = d^2_{\kcalomega}(\phi_0, \phi), $$
the square of the geodesic distance of $\kcalomega$ from the background metric (or potential)  $\phi_0$  to $\phi$. 
If instead of $dX$ we choose the Haar measure as the integration measure, it is known that the distance function $d_{\bcal_k}$ on $\bcal_k$ converges to the distance function of $\kcalomega$  \cite{CS,B}.  Thus, we formally have
\begin{equation} \label{1klog} \int_{\bcal_k} F_k(\phi) e^{- d^2_{\bcal_k}(\phi_0, \phi)}  \dcal_k\phi \to \int_{\kcalomega}  F(\phi) e^{- d_{\kcalomega}^2(\phi_0, \phi)} \dcal \phi\, ,\end{equation}
where $\dcal_k\phi$ denotes the Haar measure and $\dcal\phi$ denotes the Mabuchi measure, as usual.  In effect, by using $\frac{1}{k} \log P_k$
instead of $P_k$, the limit measure is spread out from a delta function at the background metric to a bell curve centered at the background metric.

We note that the actions do not depend a priori on derivatives of $\phi$, and thus we may expect these random potentials  to be very singular. Recall that
the corresponding metric is $\omega_0 + i \ddbar \phi$. We could smooth out the singularity by adding a term to the action involving derivatives of $\phi$, but for the moment we consider the simple case without derivative terms.

The LDP rate functional for the model \eqref{gaussmodel} was determined in
\cite{BG} and is given by
\begin{equation}
\label{GBA}
I(\nu) = g\int x^2 d\nu(x) - \Sigma(\nu) + \text{constant}\, .
\end{equation}
Here we would like to discuss the geometric meaning of this
functional in the context of \kahler geometry. The pullback of 
\eqref{GBA} to $\kcalomega$ is given by
\begin{equation} I(\phi) = g\int_M (\exp_{\phi_0}^{-1} \phi)^2 d \mu_{\phi_0}  - \int \int_{M \times M}
\log |\exp_{\phi_0}^{-1} \phi(x) - \exp_{\phi_0}^{-1} \phi (y)|
d\mu_{\phi_0}(x) d\mu_{\phi_0}(y)\, . \end{equation} 
This is new
action functional in \kahler geometry. The exponential map is not
well-defined  (see \cite{RZ}), but $\exp_{\phi_0}^{-1} \phi$ is
well-defined as the Dirichlet-to-Neumann operator, taking the
solution of the endpoint problem for geodesics to the initial data
of the geodesic. It is known that $\exp_{\phi_0}^{-1} \phi$ is
Lipschitz continuous, so the integrals are well-defined. If we
write $\dot{\phi} = \exp_{\phi_0}^{-1} \phi(x)$ for the initial
velocity, then the functional becomes
\begin{equation} I(\phi_0, \dot{\phi}_0) = g\int_M (\dot{\phi})^2 d \mu_{\phi_0}  - \int \int_{M \times M}
\log |\dot{\phi}(x) - \dot{\phi} (y)| d\mu_{\phi_0}(x)
d\mu_{\phi_0}(y). \end{equation} The first term is the one obtained in the previous section,
and is simply the Hamiltonian of the geodesic flow on $T^* \kcal_{[\omega_0]}$, which is  equal to the energy of the geodesic from $\phi_0$ to $\phi$.

%This is an important step in proving limit formulae and large
%deviations principles as $k \to \infty$. One natural method is to
%form empirical measures from the maps \begin{equation} \label{nuk}
%\nu_k: {\bf a}_k \to \mcal(\R), \;\;\; \nu_k(\mu_1, \dots,
%\mu_{N_k}) = \frac{1}{N_k} \sum_{j = 1}^{N_k} \delta_{\mu_j}
%\end{equation} from ${\bf a}_k$ to the space of probability measures on $\R$.
%Thus, we view our random variables as random measures taking
%their values in $\mcal(\R)$.  The measures $Q_k$ are then the ``pushforwards" of the measures
%$e^{- S_k} \dcal_k \phi$ under the maps (\ref{nuk}) to
%empirical measures. We work out one example in more detail for the
%sake of clarity.
%$\clubsuit$

\section{Approximation of geometric actions}

\subsection{Aubin-Yau action and the contraction principle}

In this section, we consider the top-down approach, where we construct the random sequences which converge to a given geometric functional on $\kcalomega$. We have already shown that the invariant metric  on $\bcal_k$ converges to the Mabuchi metric \eqref{MSD}, see \eqref{HaarMab}. Therefore, getting the approximation to actions is the second important ingredient in our construction. 

Here we apply the contraction principle, formulated in \S \ref{CP}, to the Aubin-Yau action \eqref{AY} and determine the corresponding sequence of finite-dimensional approximations $S_k$, which converge to $S_{AY}$ in the sense of large deviation theory. 

We claim that the minimizing potential for the variational problem
$$S_k(P) = \inf_{\{\phi \in \kcal_{[\omega_0]},\, \text{Hilb}_k(\phi) = P \in \bcal_k\}} \;\;\; S_{AY}(\phi) $$
is a Bergman metric $\phi(P) \in \bcal_k$ and that $$S_k(P) =
S_{AY}(\phi(P)) + O(1/k).$$ Thus, to leading order, the finite
dimensional approximation to the Aubin-Yau action obtained in this way is the same as its restriction to $\bcal_k$.

We set up the constrained variational problem as a
Lagrange multiplier problem. We want to find the critical
points for
$$S_{AY}(\phi) + \text{tr}\bigl( L 
(\text{Hilb}_k(\phi) - P)\bigr)\, , $$
where the Lagrange multiplier $L$ is a hermitian matrix.
We then get the following equations determining the critical point
$$
\frac1V\bigl[k+\frac12\Delta_\phi\bigr](s_\alpha L_{\alpha\beta}\bs_{\beta})h_0^ke^{-k\phi}=1\, ,\quad
\text{Hilb}_k(\phi)_{\alpha\beta}=P_{\alpha\beta}\, .
$$
Since $\Delta_\phi$ is a positive definite laplacian on $M$, the only solution to the first equation is
$$
(s_\alpha L_{\alpha\beta}\bs_{\beta})h_0^ke^{-k\phi}=V/k\, .
$$
Therefore we can solve for $\phi$ in terms of the matrix $L$,
$$
e^{-k\phi(L)}=\frac Vk\frac1{(sL\bs)h_0^k}
$$
and plug the solution to the second critical point equation. We get
$$
\frac1k\int\frac{\bs_\alpha s_\beta}{(s L \bs)}\omega_{\phi(L)}^{n}=P\, .
$$ 
The last two equations imply that $\phi(L)=FS_k(L)$, where the matrix $L$ is implicitly determined by the matrix $P$, according to the last equation. It follows that $\phi_L$ is a Fubini-Study metric, and that to
leading order
\begin{equation}
\label{AYapprox}
S_k(P) = S_{AY}\bigl(\phi(P)\bigr) + O(1/k).
\end{equation}
Therefore the contraction principle yields the same action as the pull-back of the action under the $FS_k$ map.

\subsection{Mabuchi action and Donaldson functional}

The Mabuchi action \eqref{mabuchi} has a  minimizer $\phi_c$ corresponding to the constant scalar curvature metric, which always exists in complex dimension one in each \kahler class.
The contraction principle, applied to the Mabuchi action, leads to a constrained variational problem that is very hard to solve.  

However, there exists another approximation method, proposed by Donaldson  \cite{Don2}, which is based on the pointwise Bergman kernel asymptotics \eqref{TYZ}.
The idea is to consider the variation of the $\log \det \text{Hilb}$ with respect to $\phi$,
$$
\delta \log\det \text{Hilb}_k(\phi)={\rm Tr} \, \text{Hilb}_k^{-1}(\phi)\delta  \text{Hilb}_k(\phi)=\frac1V\int_M\bigl(-k\rho_k(\omega_\phi)-\frac12\Delta_{\phi}\rho_k(\omega_\phi)\bigr)\delta\phi\,\omega_\phi^n\, ,
$$
where we have used Eq.\ \eqref{varHilb}. Using the expansion Eq.\ \eqref{TYZ}, one can then integrate the right side and get
$$
V\log\det\text{Hilb}_k(\phi)=-kN_kS_{AY}(\omega_0,\phi)+\frac k2S_M(\omega_0,\phi)+\mathcal O(1)\, ,
$$
provided that we chose the basepoint for the $\text{Hilb}$ inner product to be the identity matrix, as we've always done. Therefore the following functional on $\kcalomega$ approximates the Mabuchi action on $\kcalomega$ in the large $k$ limit
\begin{equation}\label{name}
\mathcal  L(\omega_0,\phi)=\frac {2V}k \log\det\text{Hilb}(\phi) +2N_kS_{AY}(\omega_0,\phi)\, .
\end{equation}
Following Donaldson in \cite{Don2}, let us now define the following functional on positive hermitian matrices $P\in GL(N_k,\C)/U(N_k)$,
\begin{equation}
\label{Mabapprox}
Z(P)= \frac Vk \log\det P +N_kS_{AY}\bigl(\omega_0,\phi(P)\bigr)\, .
\end{equation}
This functional is obtained by pulling back the first term in \eqref{name} from $\kcalomega$ to $\hcal_k$ using the \text{Hilb} map and by pushing forward the second term in \eqref{name} in the opposite direction using the $FS_k$ map \eqref{FSID}. The difference between the two operations is at least of order $1/k$ on $\bcal_k$, as we pointed out in \eqref{FSHilb}. Hence, the functional \eqref{Mabapprox} defines automatically a good approximation to the Mabuchi action on the space $\bcal_k$. 

We can then  consider  the following sequence of measures on positive-definite hermitian matrices in $GL(N_k,\C)/U(N_k)$,
\begin{equation}
\label{Don}
\mu_k(P)=e^{-\gamma Z(P)}d\mu_N(P)\, ,
\end{equation}
where $\gamma$ is a coupling constant and we use the Haar measure \eqref{HAAR}. 
Note that, as required, $Z(P)$ is invariant under the multiplication of $P$ by a strictly positive constant and so is the measure \eqref{HAAR}. Therefore \eqref{Don} is well-defined on the space $\bcal_k$. If we impose explicitly the constraint $\det P=1$, the functional $Z(P)$ becomes simply proportional to the functional \eqref{AYapprox}, that we obtained by applying the contraction principle to the Aubin-Yau action. The fact that this functional effectively defines the restriction of the Mabuchi functional to the space of Bergman metrics has been also observed in \cite{PS2}. 

The study of the $k\rightarrow\infty$ limit of \eqref{Don}, see also \eqref{MPI}, is a very interesting open problem that we plan to discuss in the future.

A second  approach to regularizing the Mabuchi action is to linearize the problem around the constant scalar curvature metrics by  applying the contraction principle to the modified Mabuchi functional, 
$$ \frac{1}{\epsilon}S_M(\phi_c+\sqrt{\epsilon}\,\eta)\, , $$
see section \ref{CP}.
As in the Liouville theory \cite{ZaSC}, the semiclassical expansion can be developed for the Mabuchi action around the critical point and this reduces the Mabuchi action to its gaussian approximation
\begin{equation}
\label{expansion1}
\frac1{\epsilon}S_M(\phi_c+\sqrt{\epsilon}\,\eta)= \frac18\int\eta(\Delta_0-2\bR)\Delta_0\eta\,\omega_0+ \cdots
\end{equation}
where $\Delta_0$ is the laplacian on the critical metric.
The contraction principle leads to the following constrained problem,
$$S_{M,k}(\delta P) = \inf_{D_{\phi_c} \text{Hilb}_k(\phi)  = \delta P}\;  \frac18\int\eta(\Delta_0-2\bR)\Delta_0\eta\,\omega_0\,. $$
It would be interesting to see how well this approximates our original problem.

\section{Conclusions}

Given a K\"ahler manifold $M$, we have introduced a new way to integrate over the space of K\"ahler metrics in a fixed K\"ahler class on $M$.
The main ideas of the method are:

\begin{itemize}

\item To use the \kahler potential $\phi$ as the basic dynamical variable. The corresponding metric $\omega_0 + i \ddbar \phi$ is local in $\phi$. 

\item To use \kahler quantization to construct finite dimensional approximations $\bcal_k$ to $\kcalomega$. The spaces $\bcal_k$
may be identified with symmetric spaces after fixing a background metric, hence there are many natural choices of integration measures
and of action functionals on them. There is a long developed theory of approximating geometric functionals on $\kcalomega$ by
finite dimensional ones on $\bcal_k$.

\item To use the theory of large deviations to decide when a sequence $\mu_k$ of probability measures on $\bcal_k$ tends
to a limit measure on $\kcalomega$ and to determine the limit rate function. 

\end{itemize}

Several examples were explored in some detail. Our method is very general and we believe it could be applied to and shed a new light on a wide range of problems in physics and mathematics.

\vspace{1cm}

{\bf Acknowledgments.}
We would like to thank M.\ Douglas for useful discussions.
This work is supported in part by the belgian FRFC (grant 2.4655.07), the belgian IISN (grant 4.4511.06 and 4.4514.08), the IAP Programme (Belgian Science Policy), the RFBR grants 11-01-00962 and 12-01-33071 (mol\_a\_ved), by the Ministry of Education and Science of the Russian Federation under contract 14.740.11.0081 and the NSF grant \# DMS-0904252.

\end{document}